\documentclass[%
 aip,
jap,
 amsmath,amssymb,
preprint,%
]{revtex4-1}

\usepackage{graphicx}
\usepackage{dcolumn}
\usepackage{bm}

\begin{document}


\title[Micro- and macro-world of tumors]{Bridging the gap between the micro- and the macro-world of tumors}

\author{Roberto Chignola}
\email[]{Roberto.Chignola@univr.it}
\affiliation{Dipartimento di Biotecnologie, Universit\`a di Verona \\and I.N.F.N. -- Sezione di Trieste, Strada Le Grazie 15 - CV1, I-37134 Verona, Italy}
\author{Edoardo Milotti}
\email[]{Edoardo.Milotti@ts.infn.it}
\affiliation{Dipartimento di Fisica dell'Universit\`a di Trieste \\and I.N.F.N. -- Sezione di Trieste\\Via Valerio, 2 -- I-34127 Trieste, Italy}

\date{\today}

\begin{abstract}
At present it is still  quite difficult to match the vast knowledge on the behavior of individual tumor cells  with macroscopic measurements on clinical tumors. On the modeling side, we already know how to deal with many molecular pathways and cellular events, using systems of differential equations and other modeling tools, and ideally, we should be able to extend such a mathematical description up to the level of large tumor masses. An extended model should thus help us forecast the behavior of large tumors from our basic knowledge of microscopic processes. Unfortunately, the complexity of these processes makes it very difficult -- probably impossible -- to develop comprehensive analytical models. We try to bridge the gap with a simulation program which is based on basic biochemical and biophysical processes -- thereby building an effective computational model -- and in this paper we describe its structure, endeavoring to make the description sufficiently detailed and yet understandable. 
\end{abstract}


\maketitle

\section{Introduction\label{intro}}

Physics has been remarkably successful in modeling the most basic phenomena of nature, so much so that it established a well defined procedural framework. Thus we are accustomed to start with a set of experimental facts, make conjectures, build up theories and test them, and finally make  theoretical predictions. The very same success also prompted us to think that we are somehow close to an all-embracing understanding of nature. However, when we try to apply physics to grasp the basics of biological phenomena, we are stunned by the complexity, and even eminent physicists have put forward the notion that biology probably requires some new principle. In 1989, Eugene Wigner said that ``[It is] important to realize that physics does not contain the theory that I live and have desires and emotions. \dots The perfect science will give a reasonable description of the fact that we have emotions and desires, but present-day physics denies the existence of emotions and desires. It cannot be denied in my opinion.''\cite{Wigner1989} Erwin Schr\"odinger instead wrote in the first chapter of his precious book {\it What is Life?}, that ``The large and important and very much discussed question is: How can the events in space and time which take place within the spatial boundary of a living organism be accounted for by physics and chemistry? The preliminary answer which this little book will endeavor to expound and establish can be summarized as follows: The obvious inability of present-day physics and chemistry to account for such events is no reason at all for doubting that they can be accounted for by those sciences.''\cite{schrodinger1944life} Thus Schr\"odinger was more hopeful, and indeed, measurements in biology display the same regularity to which we are accustomed in the study of physical phenomena, and at least some biophysical and biochemical phenomena are by now quite well understood in terms of basic physical laws. So what is missing from the big picture? We believe that the answer lies in the fog of complexity. Biological phenomena are vastly more complex than many physical phenomena, and this prevents us from using many time-tested modeling techniques. In particular, it is very difficult to step from the descriptions of molecular pathways in cells up to the behavior of whole organs, not to mention something which is even more complex like ``emotions and desires''. 

The remarks by Wigner and Schr\"odinger set the ultimate goals of any physical theory of life, however we are still very far away from it, and we must lay the humble groundwork first. Interestingly, tumors provide both a good starting point and a testing ground for computational models, because tumor cells lack some of the control paths of normal cells and tumors display clear and well defined growth patterns \cite{hanahan2000hallmark}, which are evidently amenable to mathematical modeling. Tumors also set an immediate, very practical goal, that readily obscures the philosophical underpinnings of this kind of research, because mathematical models of cancer growth can lead to a better understanding of tumor therapy.

For instance, many solid tumors fit quite well an old phenomenological growth law, the {\it Gompertz law}, which is defined, in the case of volume, by the single two-parameter formula 
\begin{equation}
V(t) = V(0) \exp\left[ \frac{\alpha_G}{\beta_G} \left( 1-e^{-\beta_G t} \right) \right]
\end{equation}
where $V$ is the tumor volume, and $\alpha_G$ and $\beta_G$ are numerical fit parameters. The simplicity and effectiveness of this law are  enticing, because they seem to indicate that with some effort we could try to use this information to reach a better understanding of the inner machinery of tumors. Unfortunately it is not so, and while much work has gone in the attempt to find a fundamental model of the Gompertz law that may actually connect it to the microscopic clockwork of tumor cells, nobody has succeded to this day\cite{bajzer2000new}.

In the case of tumors, modeling efforts are by no means limited to the simple Gompertz law: many researchers have attacked the problem of tumor modeling from different points of view (for a review see, e.g., Ref. \onlinecite{tracqui2009biophysical}), and with a broad range of mathematical and numerical techniques. Remarkably, different approaches yield appealing qualitative results, but is this a measure of success? Again, we believe that a really successful model should go beyond a seeming resemblance (biomimesis), and bridge the gap between the microscopic description of tumor biology -- which is by now extremely well developed thanks to the progress of molecular biology and all the ``-omic" disciplines -- and the macroscopic measurements. Indeed, the connection between micro- and macro-world is not an end in itself: besides its great importance for our understanding of the basics of life, it is also of great practical value, because it would help managing illness and create new drugs (for similar reflections on this theme, see Ref. \onlinecite{Hunt:2009:Pharm-Res:19756975}).

One of the difficulties of model building is the inclusion of the discrete events that mark the life of individual cells, like the transitions between cell phases and cell division. This is often skipped over, assuming that the discreteness is washed away by some sort of averaging. However this is certainly not true at the very earliest stages of tumor development, when cells are few, and in general it is not valid because tumors are multiscale phenomena, where microscopic and macroscopic scales are intimately linked by diffusion in the microenviroment\cite{Nitsche1999}. For instance, cell synchronization and other such collective behaviors which are determined by microscopic events could be  important in future approaches to tumor therapy\cite{heinemann2010effect}, and therefore it is important not to lose sight of these discrete, microscopic events even in large tumor masses. Then, the need to include discreteness means that detailed analytical models -- that assume a continuous time evolution, and often also neglect the fine-grained structure at the cellular level replacing it with continuous space-dependent functions -- are ruled out. 

This leaves us with the numerical tools. However the most direct way, the ground-up simulation that starts from atoms and molecules and reaches up to the level of large multicellular clusters is also impracticable. It is easy to see that a simulation that uses the methods of standard molecular dynamics is utterly unfeasible\cite{chignola2004numerical,chignola2010}. Here we describe how we confront this problem developing a simulation program of tumor spheroids (an {\it in vitro} model of solid, unvascularized solid tumors, see, e.g., Ref. \onlinecite{sutherland1988cell}) that uses an in-between solution, and works at the single-cell level. 

\section{Overall structure of the simulation program \label{struct} }

While we already described the present version of the program in full detail in Ref.~\onlinecite{milotti2010emergent}, in this section we provide a brief summary discussion of its structure.  
The main logic blocks of the simulation program are as follows (see also figure  \ref{layout}):

\begin{figure}[!ht]
\begin{center}
\includegraphics[width=\linewidth]{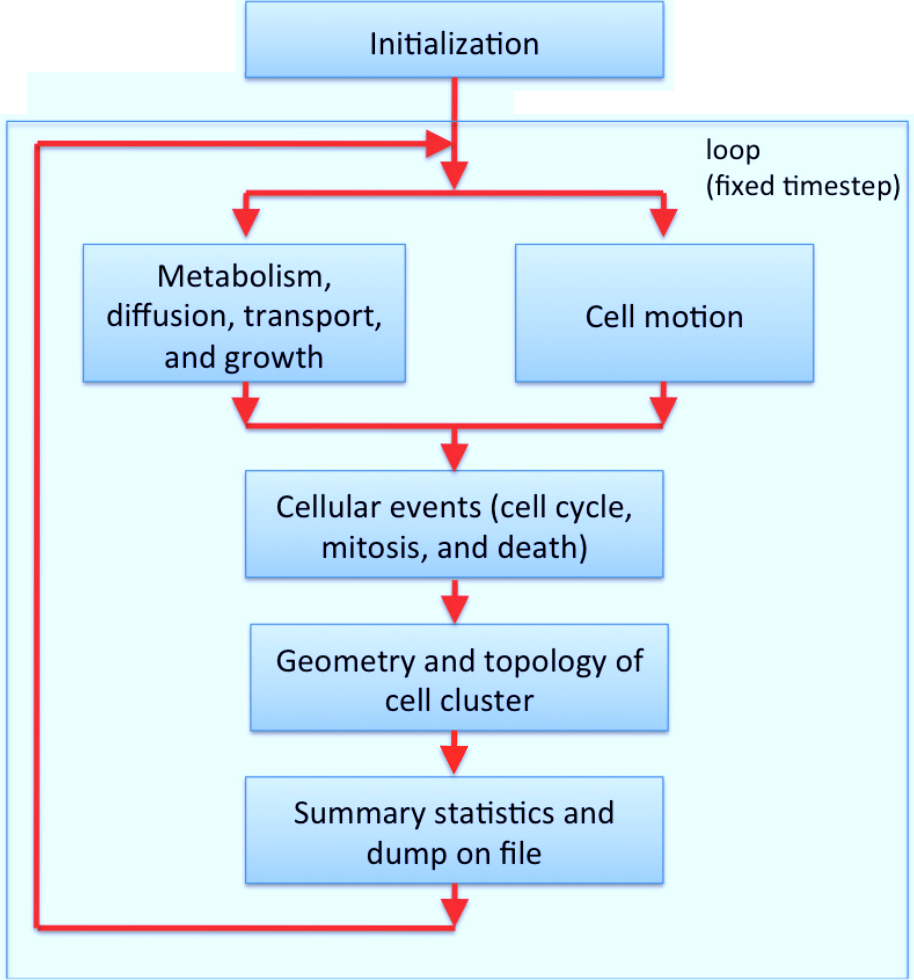}
\end{center}
\caption{Block diagram of the simulation program, reproduced from Ref. \onlinecite{milotti2010emergent}.}
\label{layout}
\end{figure}

 \begin{description}
 \item[Initialization] this part of the program defines the kind of simulation, like the number of initial cells and the initial environment. Moreover, during this step, cells are allowed to run free for a large number of cell cycles, while proliferation and death are frozen, so that metabolic values approach an initial equilibrium. 
 \item[Metabolism, diffusion, transport and growth] metabolism and diffusion are deeply intertwined in the mathematical description, and the basic equations correspond to a discretized diffusion-reaction problem; this is discussed in detail for one chemical species -- glucose -- in section \ref{glucbio}, and during this step the procedure is repeated for all substances included in the simulation; cell volume is also a dynamical variable linked to cell biochemistry, and therefore this step computes cell growth as well as the concentrations of different chemical species in cells and extracellular spaces. 
 \item[Cell motion] as cells grow and proliferate, they push against neighboring cells, and the cell cluster must rearrange its shape; the biomechanical evolution of the cell cluster is described in section \ref{biomech} below, and can be executed in parallel with the metabolic evolution. In practice this is implemented with multithreading in an OpenMP environment\cite{dagum1998openmp}.
 \item[Cellular events] in addition to the smooth biochemical and biomechanical evolution, cells also undergo sudden changes of state, activated by internal biochemical switches. The simulation program includes some important checkpoint, which are modeled with different accuracies, mitosis, and several conditions for cell death\cite{Chignola2005,chignola2007ab}. Some of the events are partly stochastic, e.g., at mitosis, organelles are shared between daughter cells according to a binomial distribution\cite{catlett2000divide,bergeland2001mitotic}.
 \item[Geometry and topology] to compute both diffusion and cell-cell forces we must know the proximity relations between cells. This is done in a specialized part of the program, which uses the computational geometry library CGAL to calculate the Delaunay triangulation and the alpha-shape of the cell cluster\cite{cgal}. The Delaunay triangulation\cite{de2008computational}  returns the proximity relations, while the alpha-shape\cite{edelsbrunner1992three,dey1999computational} defines the boundary between the cluster and the outer environment.
 \item[Summary statistics] the program produces a huge amount of simulated data, and this part deals both with the output of summary statistics, and with periodic output of raw data to file. The data sent to file are analyzed offline with specialized analysis programs. This final block is not essential to the whole machinery that drives the simulation, but is all-important to assess the results.
 \end{description}

\section{Computing the biochemistry of glucose \label{glucbio}}

\begin{figure}[!ht]
\begin{center}
\includegraphics[width=\linewidth]{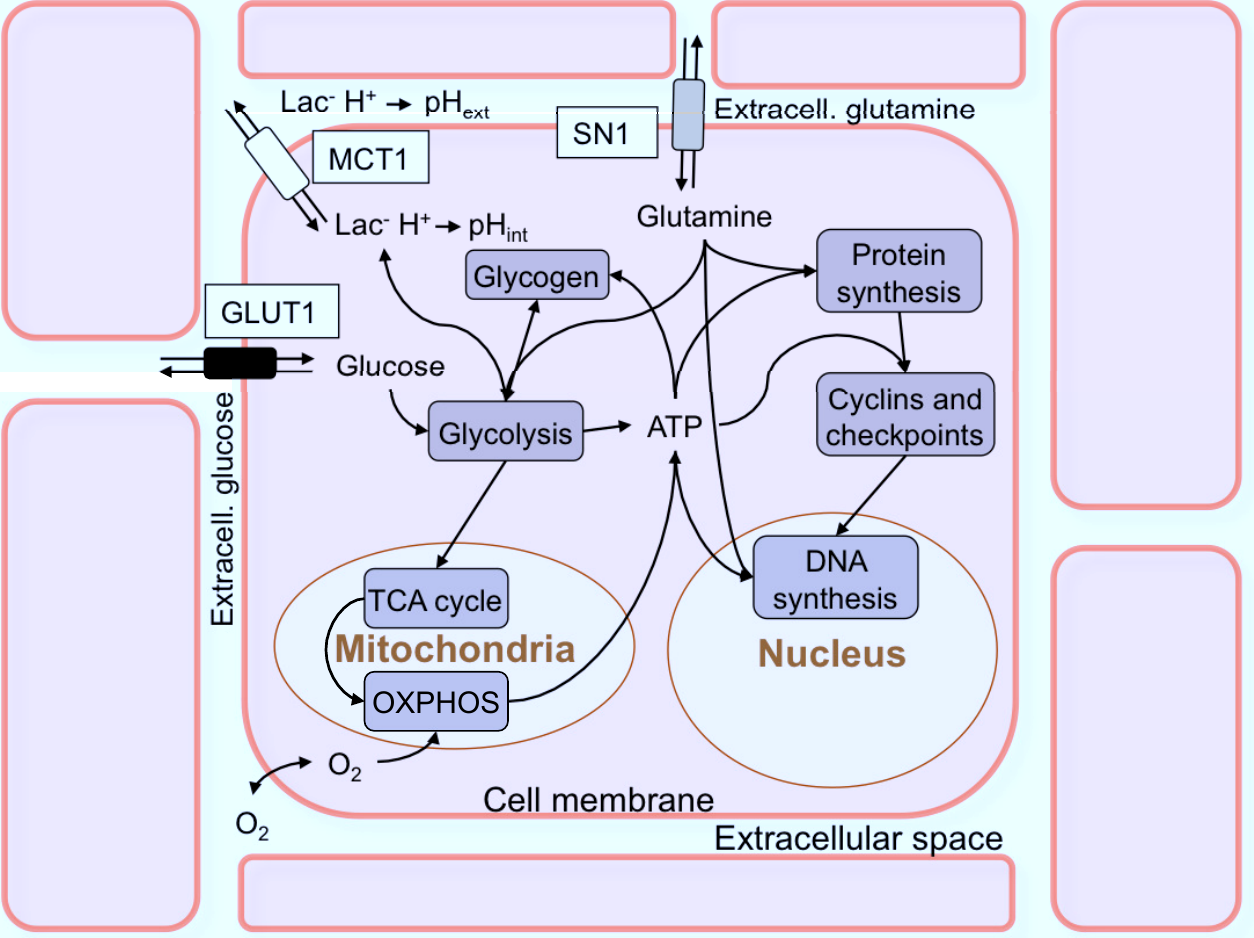}
\end{center}
\caption{Schematic diagram of our simplified metabolic network, reproduced from Ref. \onlinecite{milotti2010emergent}. For additional details see Refs. \onlinecite{milotti2010emergent,Chignola2005,chignola2007ab}.}
\label{metnet}
\end{figure}

Tumor cells behave as small automata that  proliferate all the time, and  it is extremely important to model the proliferative machinery as well as possible, from the metabolic network that provides energy and nutrients, to the biomechanics of the resulting cell cluster. These aspects actually influence each other, as the structure of the cell cluster determines the diffusion of different chemical species, and therefore also the transport and reaction processes of individual cells. Now it is important to recall that diffusion processes in cell clusters are often mediated by molecules on the cells' membranes. These facilitated diffusion processes need the extracellular spaces to proceed, and therefore we model cells as two-compartment objects. The inner compartment is the cell proper, while the outer compartment is the  extracellular space around the cell. Each cell communicates with its extracellular space only, while adjacent extracellular spaces exchange metabolites and nutrients that are transported into and out of cells by facilitated diffusion. Figure \ref{metnet} shows a scheme of the simplified metabolic network that we implemented in our description of individual cells. 

Now, to exemplify the actual implementation of this scheme, we concentrate our attention on glucose, one of the principal molecules involved in the metabolic network. The equations that regulate the transport and metabolism of glucose in the simulation program  are
\begin{subequations}
\begin{eqnarray}
\label{gdiff}
\frac{d{{m}_{G,c}}}{dt} &=& {{D}_{G}}\sum\limits_{\left\langle b \right\rangle }{\left( \frac{{{m}_{G,b}}\left( t \right)}{{{V}_{b}}\left( t \right)}-\frac{{{m}_{G,c}}\left( t \right)}{{{V}_{c}}\left( t \right)} \right){{g}_{bc}}}-{{T}_{G}}\left[ {{m}_{G,c}}\left( t \right),{{m}_{G,C}}\left( t \right) \right] \\
\label{gmet}
\frac{{d{m_{G,C}}}}{{dt}} &=& {M_G}\left[ {{m_{G,C}}\left( t \right)} \right] + {T_G}\left[ {{m_{G,c}}\left( t \right),{m_{G,C}}\left( t \right)} \right]
\end{eqnarray}
\end{subequations}
where we note first of all that $c$ is the same index as $C$ -- uppercase denotes the cell and lowercase denotes the corresponding extracellular space -- and $\langle b \rangle$ denotes the  set of indices of the extracellular spaces adjacent to $c$. This set $\langle b \rangle$ is continually changing since the model is lattice-free,  and the adjacency relationships among cells must be computed step by step as the structure evolves under the mechanical pushes of cell volume growth and of the occasional mitoses (cell divisions). Moreover $m_{G,C}$ is the mass of glucose inside  cell $C$, ${{m}_{G,c}}$ is the mass of glucose in the surrounding extracellular space, $D_G$ is the diffusion coefficient of glucose in the extracellular space (this takes into account the complex structure and composition of the extracellular spaces), $V_c$ is the volume of the extracellular space, $g_{bc}$ is a geometric factor\cite{milotti2010emergent} that takes into account the distance and the contact surface between extracellular spaces $b$ and $c$,  $T_G$ is a (facilitated) transport term that depends primarily on the concentrations (and therefore on the masses) of glucose inside the cell and in the extracellular space, and finally $M_G$ is a term that describes glucose metabolism inside the cell (and depends primarily on glucose mass inside the cell). 

The first term on the  r.h.s.  of equation ({\ref{gdiff}) corresponds to a discretization of glucose diffusion in the extracellular spaces \cite{milotti2010emergent}.
The transport and the metabolic terms in equations (\ref{gdiff}-\ref{gmet}) are defined by the following formulas:
\begin{subequations}
\begin{eqnarray}
\label{TG}
{T_G} & = & a2c\cdot\frac{{{v_{\max ,1}}{m_{G,c}}}}{{{V_c}{K_1} + {m_{G,c}}}} - c2a \cdot \frac{{{v_{\max ,1}}{m_{G,C}}}}{{{V_C}{K_1} + {m_{G,C}}}} \\
\label{MG}
{M_G} & = &  - \frac{{{v_{\max ,2}}{m^2}_{G,C}}}{{\left( {{V_C}{K_2} + {m_{G,C}}} \right)\left( {{V_C}{K_a} + {m_{G,C}}} \right)}} - \frac{{{v_{\max ,22}}{m^2}_{G,C}}}{{\left( {{V_C}{K_{22}} + {m_{G,C}}} \right)\left( {{V_C}{K_a} + {m_{G,C}}} \right)}}
\end{eqnarray}
\end{subequations}
The two double Michaelis-Menten terms in the metabolic function (\ref{MG}) correspond to the glucose transformation into glucose-6-phosphate, a molecule that remains trapped inside the cell, due to hexokinase and glucokinase activity \cite{Chignola2005}, where the minus sign means that this decreases the total glucose mass, while the simple Michaelis-Menten terms in the transport function (\ref{TG}) represent transport into (first term, positive) and out of the cell (second term, negative). Several parameters in these equations have their usual meaning, e.g., the various $v_{max}$ are the maximum rates at which the Michaelis-Menten reactions proceed, and the $K$'s are the corresponding k-values. The transport coefficients $a2c$ and $c2a$,  may seem to be redundant, but they have been introduced to parameterize corrections to transport due to environmental variables such as acidity, and they are defined as follows to fit the measured transport efficiency \cite{Chignola2005,chignola2007ab}:
\begin{subequations}
\begin{eqnarray}
a2c &=& \frac{1}{2}\left[ {1 + \tanh \left( {\mathrm{a2c\_slope}\cdot pH_c - \mathrm{a2c\_thr}} \right)} \right] \\
c2a &=& \frac{1}{2}\left[ {1 + \tanh \left( {\mathrm{c2a\_slope}\cdot pH_C - \mathrm{c2a\_thr}} \right)} \right]
\end{eqnarray}
\end{subequations}
where $pH_C$ and $pH_c$ are, respectively, the internal pH and the pH in the extracellular space. The dependence of glucose transport on the cell's surface area $S$ is encoded in the definition of $v_{\max ,1}$ as well (see Refs. \onlinecite{Chignola2005,chignola2007ab}): 
\begin{equation}
{v_{\max ,1}} = \mathrm{VMAX1}\cdot h \cdot S
\end{equation}
where $h$ is yet another corrective factor\cite{Chignola2005,chignola2007ab}, 
\begin{eqnarray}
\nonumber
h &=& 0.5\left[ {1.3\left( {1 - \frac{{{m_{O2,C}}}}{{{V_C}\;\mathrm{O2st}}}} \right) + 1} \right]\cdot\left\{ {1 + \tanh \left[ {100\left( {1 - \frac{{{m_{O2,C}}}}{{{V_C}\;\mathrm{O2st}}}} \right)} \right]} \right\} \\
&& + 0.5\left\{ {1 - \tanh \left[ {100\left( {1 - \frac{{{m_{O2,C}}}}{{{V_C}\;\mathrm{O2st}}}} \right)} \right]} \right\}
\end{eqnarray}
related to the oxygen mass inside the cell, $m_{O2,C}$. These expressions are parameterizations of observed pH- and oxygen-dependent transport activity, and the related parameters are $\mathrm{VMAX1}$, $\mathrm{a2c\_slope}$, $\mathrm{a2c\_thr}$, $\mathrm{c2a\_slope}$, $\mathrm{c2a\_thr}$, and $\mathrm{O2st}$; we have introduced hyperbolic tangents as a practical way to avoid the sharp kinks associated to the spline functions quoted in the literature\cite{Chignola2005,chignola2007ab}, and at the same time keep the same general shape.

The cells that are in direct contact with the environment require a slightly different form of the transport and diffusion equation:
\begin{equation}
\label{gdiffenv}
\frac{{d{m_{G,c}}}}{{dt}} = {D_G}\sum\limits_{\left\langle b \right\rangle } {\left( {\frac{{{m_{G,b}}\left( t \right)}}{{{V_b}\left( t \right)}} - \frac{{{m_{G,c}}\left( t \right)}}{{{V_c}\left( t \right)}}} \right){g_{bc}}}  - {T_G}\left[ {{m_{G,c}}\left( t \right),{m_{G,C}}\left( t \right)} \right] + {D_{G,env}}\left( {{\rho _{G,env}} - \frac{{{m_{G,c}}}}{{{V_c}}}} \right){g_c}
\end{equation}
where $\rho _{G,env}$ is the glucose concentration in the environment, $g_c$ is the geometric factor between cell and environment, and $D_{G,env}$ is the diffusion coefficient of glucose in the environment.  We must also add the equation for the environmental glucose itself:
\begin{equation}
\label{genv}
\frac{{d{\rho _{G,env}}}}{{dt}} =  - \frac{1}{{{V_{env}}}}\left( {\sum\limits_{\left\langle c \right\rangle } {{D_{G,env}}\left( {{\rho _{G,env}} - \frac{{{m_{G,c}}}}{{{V_c}}}} \right){g_c}} } \right) + \left( {{\rho _{G,in}} - {\rho _{G,env}}} \right)\frac{f}{{{V_{env}}}}
\end{equation}
where $V_{env}$ is the volume of the environment,  $f$ is the flow of nourishing solution into the environment (there is a corresponding outflow of spent solution), and $\rho _{G,in}$ is the glucose concentration in the nourishing solution.

At each time step the program must find the updated values of glucose mass both inside cells and in the extracellular spaces, and it does so by solving equations (\ref{gdiff}-\ref{gmet}), (\ref{gdiffenv}) and (\ref{genv}) for both cells and environment. Now it is important to note that glucose is just one of several molecules that the program tracks at each time step, and we have seen above that there are several ways in which these molecules actually interact, e.g., glucose transforms into glucose-6-phosphate and the glucose transport coefficients are influenced by the pH of the surrounding extracellular spaces: this means that the equations of the molecular species are all coupled. Moreover, there are many such equations: the present model includes 19 equations per cell, and since the number of cells can be of the order of one million, the program must eventually be able to solve tens of millions of coupled nonlinear differential equations. At first sight this looks like a formidable task, however we are helped here by some fortunate conditions. Firstly, the equations of metabolism lead to biological homeostasis, i.e., the complete dynamical system is dissipative and has at least a stable fixed point. This means that a simple integration method like implicit Euler works perfectly well for this kind of differential equations\cite{milotti2009numerical}. Moreover, implicit methods invariably lead to systems of nonlinear equations, and here the resulting system is just as large as the original system of differential equations, however if we take time steps that are sufficiently small (and we shall see later how small they have to be) then the solution of the system is actually quite close to the solution at the previous time step, so that even a conceptually simple method like inexact Newton-Raphson converges to the new solution in a reasonable CPU time. A more detailed description of the numerical procedure can be found in Refs. \onlinecite{milotti2009numerical,milotti2010emergent}.

\section{Biomechanics \label{biomech}}

Just as the metabolic processes proceed, cells grow, proliferate and die, and this leads to internal movements and structural changes in  clusters of cells. This biomechanical part is exceedingly complex and in our simulation program we must resort to  simplifications and phenomenological parameterizations. The first and foremost simplification is that cells are treated almost everywhere as simple spheres, so that four numbers suffice to specify their position and size. Then, forces between cells are computed as functions of the distance of their centers and of their radii.

We use a parameterization of the cell-cell forces that is similar to that used in Refs. \onlinecite{Palsson:2001fk,Galbusera:2008}. The basic idea is that for small deformations of the cell membrane, it is reasonable to assume that it behaves as the elastic membranes of the Hertz problem (interaction of two spherical membranes) or as in the Boussinesq problem (axisymmetric pressure on a flat membrane)\cite{Wei:2001,Galle:2005}. In both cases the force is proportional to  ${k_C}\left|x\right|^{3/2}$, where $x$ is the relative deviation from the equilibrium position and ${k_C}$ is a constant related to the problem parameters. In the Hertz problem we find that\cite{milotti2010emergent} 
\begin{equation}
{k_C} = \frac{{\sqrt {{R_1}{R_2}} \left( {{R_1} + {R_2}} \right)}}{{\frac{3}{4}\left( {\frac{{1 - \nu _1^2}}{{{E_1}}} + \frac{{1 - \nu _2^2}}{{{E_2}}}} \right)}} 
\end{equation}
with
\begin{equation}
x = \frac{{d - \left( {{R_1} + {R_2}} \right)}}{{\left( {{R_1} + {R_2}} \right)}}
\end{equation}
and where $R_{1,2}$ are the radii of the two interacting spheres, $d$ is the distance between the centers, $E_{1,2}$ is Young's modulus and $\nu_{1,2}$ the Poisson ratio of each sphere.

Under compression the resulting repulsive force can be quite large, especially after cell division (mitosis): such large forces are not observed in real cells, and thus we assume that at small enough separation $d$, the force flattens out and has a constant modulus. We set the position of this flattening so that we obtain the observed duration of mitosis.

When cells move apart the force is attractive, because of adhesion molecules on the cell membrane\cite{Zhu:2000:J-Biomech:10609515}. The force range of the adhesion molecules is quite small (tens of nanometers) but here we assume a far larger range, as large as a few microns: in this way we account -- albeit phenomenologically -- for shape deformations in the case of attractive forces. 

Moreover, cell-cell adhesion depends on the number of links between adhesion molecules on both cell membranes \cite{Bell:1984:Biophys-J:6743742}, and has a stochastic nature. If we assume a roughly Gaussian probability density which depends on the relative deviation $x$ from the equilibrium position, then 
\begin{equation}
p\left( x \right) = \frac{1}{{\sqrt {2\pi {\sigma ^2}} }}\exp \left( { - \frac{{{{\left( {x - {x_0}} \right)}^2}}}{{2{\sigma ^2}}}} \right)
\end{equation}
is the probability density that a link is detached at relative distance $x$, where $x_0$ and $\sigma$  are parameters that must be adjusted. This means that the average number of detached links at relative distance $x$ is proportional to the cumulative probability
\begin{equation}
\label{pform}
P\left( x \right) = \int\limits_0^x {p\left( {x'} \right)dx'}  = \frac{1}{2}\left[ {1 + {\mathop{\rm erf}\nolimits} \left( {\frac{{x - {x_0}}}{{\sqrt 2 \sigma }}} \right)} \right]
\end{equation}
Computing the error function is time-consuming, therefore we replace equation (\ref{pform}) with the approximate expression 
\begin{equation}
P\left( x \right) \approx \frac{1}{2}\left[ {1 + \tanh \left( {\sqrt {\frac{2}{{\pi {\sigma ^2}}}} \left( {x - {x_0}} \right)} \right)} \right]
\end{equation}
which allows a faster evaluation and approximates equation (\ref{pform}) everywhere to better than 2\%.

Finally we assume a force that is proportional both to the number of links and to the previously calculated Hertz (or Boussinesq) behavior
\begin{equation}
{F_a}\left( x \right) \approx \frac{1}{2}\left[ {1 + \tanh \left( {\sqrt {\frac{2}{{\pi {\sigma ^2}}}} \left( {x - {x_0}} \right)} \right)} \right]{k_C}{\left| x \right|^{{3 \mathord{\left/
 {\vphantom {3 2}} \right.
 \kern-\nulldelimiterspace} 2}}}
\end{equation}

Although cell-cell forces are quite important, the development of a cell cluster is propelled by the growth of individual cells, by their proliferation, and by the shrinking that takes place after cell death. Growth itself is driven by the metabolic processes, and moreover cellular volume is a notoriously complicated variable \cite{lang1997regulating,wehner2003cell}, which is related to osmotic pressure, and thus to the concentration of many substances inside cells. In our approach we take the total ATP mass as representative of this vast array of substances, and we assume that the cellular volume is partly determined by the total ATP mass. In addition, there are fixed volume contributions from the cell nucleus and from the organelles. The organelles themselves are variable in number, and again, we take the number of mitochondria as representative of the whole class of cellular organelles. These considerations yield the following phenomenological formula for the volume of living cells: 
\begin{equation}
V_C = {V_{min }}\left( {1 + \mathrm{DNA}} \right) + \mathrm{C2\cdot Mit + C1\cdot ATPp}
\end{equation}
where $V_{min}$ is the nuclear volume, related to the DNA content of the cell, DNA is the fraction of newly synthesized DNA, Mit is the number of mitochondria, ATPp is the ATP mass in the cell, and finally C1 and C2 are phenomenological parameters (parameter values are given in Ref. \onlinecite{milotti2010emergent}).

Mitosis accounts for additional strong localized pushes: here is how it proceeds in the simulation program. The metabolic network implemented in the program sets the cell's biological clock, and thanks to the inclusion of several checkpoints the program determines the entry into each main phase\cite{milotti2010emergent}. When a cell finally completes the M-phase (where M stands for mitiosis) it is replaced by two daughter cells. Then, the program selects a random direction for the axis that joins the centers of the daughter cells, and it computes the new positions of their centers. Since the total volume of the daughter cells is equal to the volume of the mother, the radius of each daughter cell is roughly equal to 80\% of the radius $R_0$ of the mother cell, i.e., the distance between the new centers is about $0.4 R_0$ (see figure \ref{mit}). Again, we remark that the centers are only representative of an ``average'' cell position and are used to compute forces; the new cells are actually compressed and deformed to fit in the original volume. 
The distance $0.4 R_0$ also sets the maximum value of the repulsive force: indeed when the daughter cells are not surrounded by other cells, the distance of their centers must increase from about $0.4 R_0$ to $1.6 R_0$ in a time equal to the duration of the M-phase. For mother cells with radius $R_0 \approx 5 \mu\mathrm{m}$ this means that the distance traveled by each cell is about $3 \mu\mathrm{m}$. Finally, if we take the total duration of the M-phase about 2000 s, we find that  the average speed $v$ of each cell is about 1.5 nm/s, and the average repulsive force is $F = \gamma v \approx $ 30 pN, taking a cell-cell viscosity $\gamma \approx 200 \mathrm{Pa}\cdot\mathrm{s}$ (see Ref. \onlinecite{milotti2010emergent} for a discussion of cell-cell viscosity). In this way we set the maximum repulsive force in the previous model of cell-cell force derived from membrane-membrane interaction.

\begin{figure}[!ht]
\begin{center}
\includegraphics[width=\linewidth]{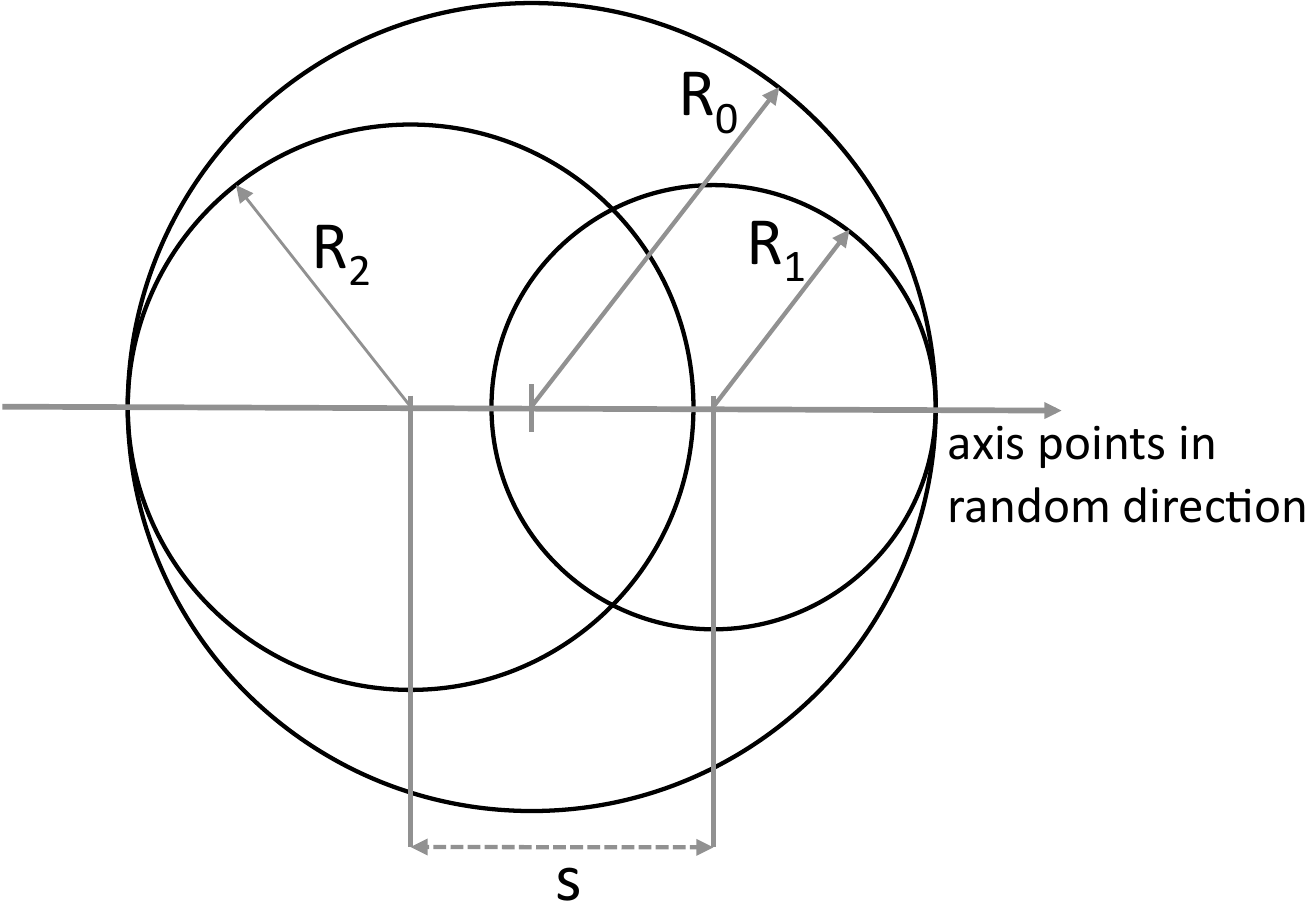}
\end{center}
\caption{Geometry of mitosis, reproduced from Ref.  \onlinecite{milotti2010emergent}. For additional details see the main text and  Ref. \onlinecite{milotti2010emergent}.}
\label{mit}
\end{figure}

Finally, a cell can die -- because of poor metabolism, or because of other external conditions, like irradiation -- and a cell cluster then feels an additional pull, as the volume of dead cells gradually shrinks\cite{bortner2002apoptotic} according to the equation
\begin{equation}
\frac{{d{V_C}}}{{dt}} =  - \mathrm{DVap}\cdot{V_C},
\end{equation}
which has the exact solution 
\begin{equation}
{V_C} = {V_0}\exp \left( { - \mathrm{DVap}\cdot t} \right),
\end{equation}
where DVap is yet another cellular parameter, and $V_0$ is the cell's volume at death\cite{milotti2010emergent}.

All  cellular motions take place in highly viscous environment. The high viscosity implies that Brownian motion in the cell cluster is negligible\cite{milotti2010emergent}, thus we use dynamical equations for the motion of cells which are similar to those of dissipative particle dynamics\cite{flekky2000foundations,dzwinel2002mesoscopic}:
\begin{equation}
{m_n}\frac{{d{{\bf{v}}_n}}}{{dt}} =  - \gamma {{\bf{v}}_n} - \sum\limits_k {{\gamma _{n,k}}\frac{{\left( {{{\bf{v}}_n} - {{\bf{v}}_k}} \right)\cdot\left( {{{\bf{r}}_n} - {{\bf{r}}_k}} \right)}}{{{{\left| {{{\bf{r}}_n} - {{\bf{r}}_k}} \right|}^2}}}\left( {{{\bf{r}}_n} - {{\bf{r}}_k}} \right)}  + {{\bf{F}}_n} + \sum\limits_k {{{\bf{F}}_{n,k}}} 
\end{equation}
where the indices $n$ and $k$ denote cells, the sums $\sum_k$ are over all neighboring cells, $m_n$ denotes the mass of the $n$-th cell, $\mathbf{r}_n$ and $\mathbf{v}_n$ are position and velocity vectors of the $n$-th cell, $\gamma$ is the friction coefficient associated with the environment (which is mostly the surrounding extracellular space), $\gamma_{n,k}$ is the friction between the $n$-th and the $k$-th cell, $F_n$ is an external force (gravity), and finally $F_{n,k}$ is the force that cell $k$ exerts on cell $n$. Finally we find a system of coupled nonlinear differential equations that describes the dynamics of the whole cluster of cells, and is loosely coupled with the biochemical part of those processes that link the mechanical behavior of cells (volume growth, mitosis, shrinking after death) with their internal biochemistry. The biomechanical equations can be solved with implicit methods like those used in the biochemical part\cite{milotti2010emergent}.
  
\section{Visualizing the  results of simulation runs \label{vis} }

Although hard data are often best represented by histograms, graphs, and other traditional plots, here it is just as  important to produce pictures that link the simulation results with the laboratory experience of biologists. 

As it happens with other complex simulations, visualization is an exceedingly important part of whole project, and at the moment we use specialized programs written in the language of {\it Mathematica}\cite{Mathematica8}, and Paraview scripts\cite{paraview} to visualize data. 
We believe that these routines can be much improved, both in quality and in computational efficiency, however they are already developed to the point where one can glimpse the power of visualization, and figures \ref{large3D}-\ref{small-4} provide a gallery of examples. 

We  also add a couple of more conventional plots that hint at some interesting, potentially useful, effect (figure \ref{small-5}). 

\begin{figure}[!ht]
\begin{center}
\includegraphics[width=\linewidth]{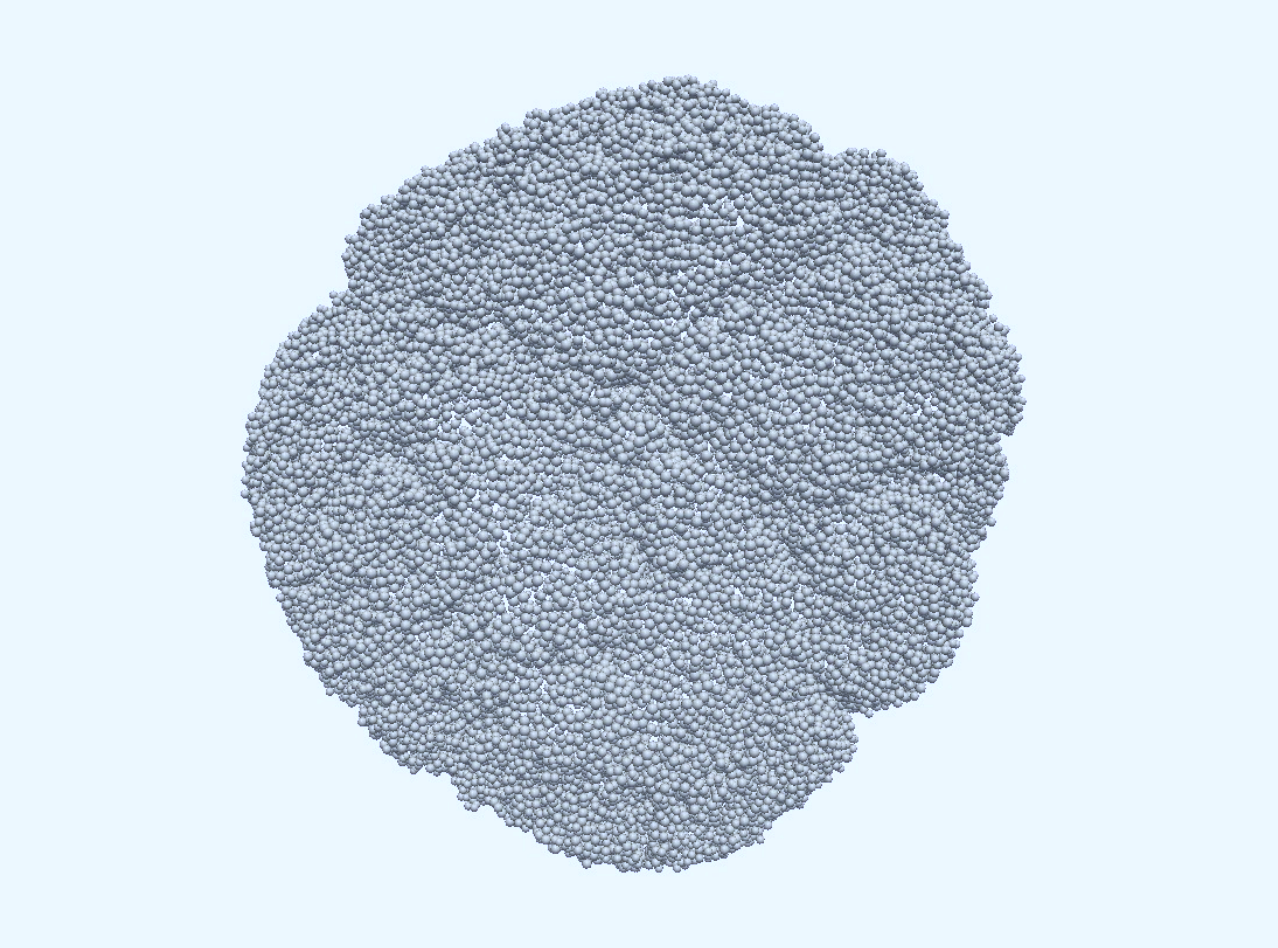}
\end{center}
\caption{Alpha-shape \cite{edelsbrunner1992three} of a large simulated tumor spheroid, displayed using ParaView\cite{paraview}. The cells shown are only those on the boundary with the environment, and the occasional white spaces mark the spots where the background shows through the closely packed structure. This simulation has been carried out with weak adhesive forces, and the spheroid deviates markedly from a spherical symmetry. On the whole the pictured spheroid contains more than half a million cells, and the approximate diameter of the spheroid is almost 800 $\mu$m. The simulation started from a single cell, and the simulated time corresponds to more than 23 days. This figure can be compared with pictures of actual tumor spheroids (see, e.g., Ref. \onlinecite{yu2003holographic,yu2004holographic}).}
\label{large3D}
\end{figure}

\begin{figure}[!ht]
\begin{center}
\includegraphics[width=\linewidth]{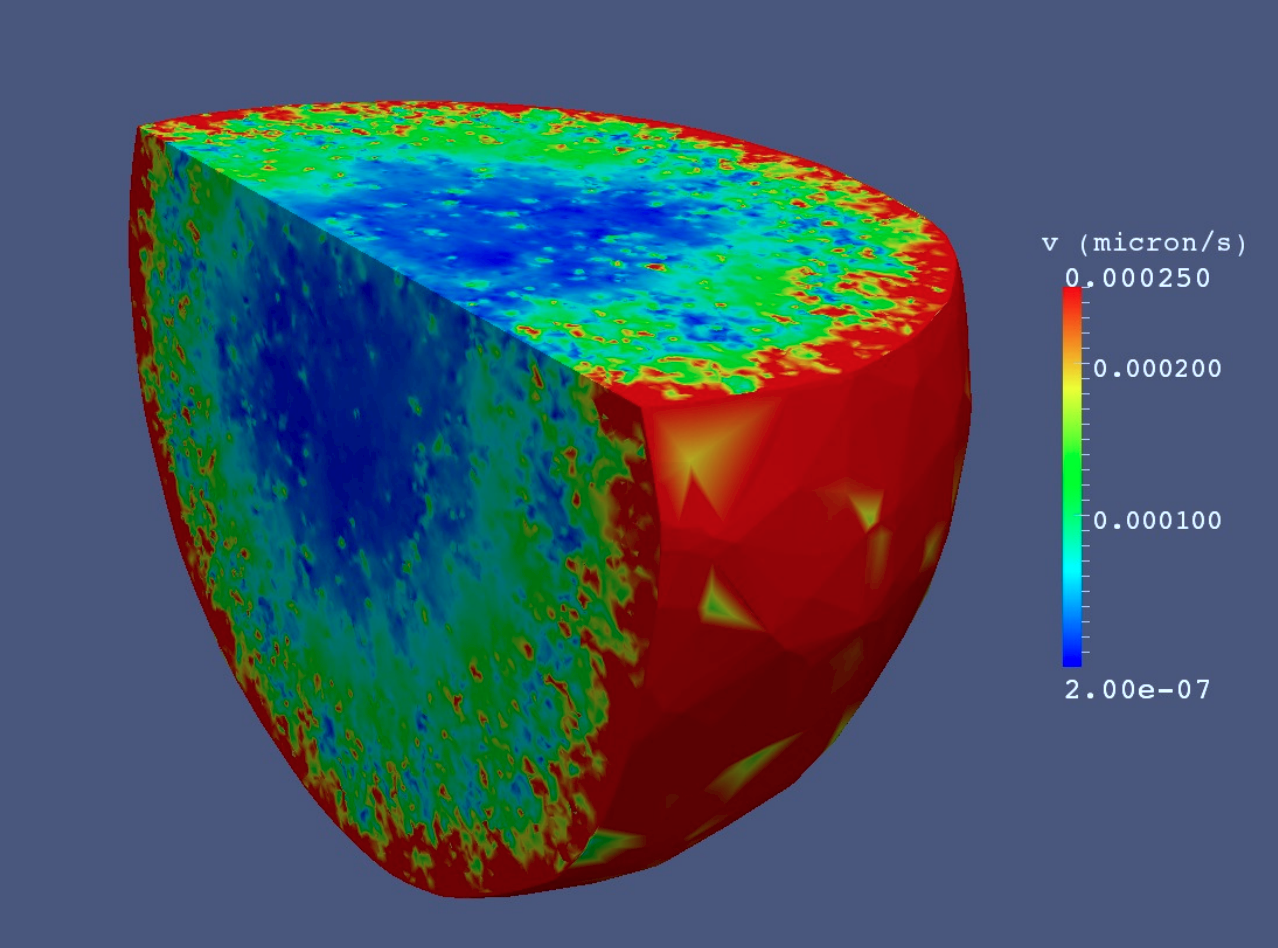}
\end{center}
\caption{Clipped map of the magnitude of the velocity field in a large tumor spheroid (about 1.2 million cells), displayed using ParaView\cite{paraview}. This map is very similar to those found in actual measurements\cite{nolte2011tissue}. }
\label{large3D-Vclipped}
\end{figure}

\begin{figure}[!ht]
\begin{center}
\resizebox{5in}{!}{%
$\begin{array}{c}
\includegraphics{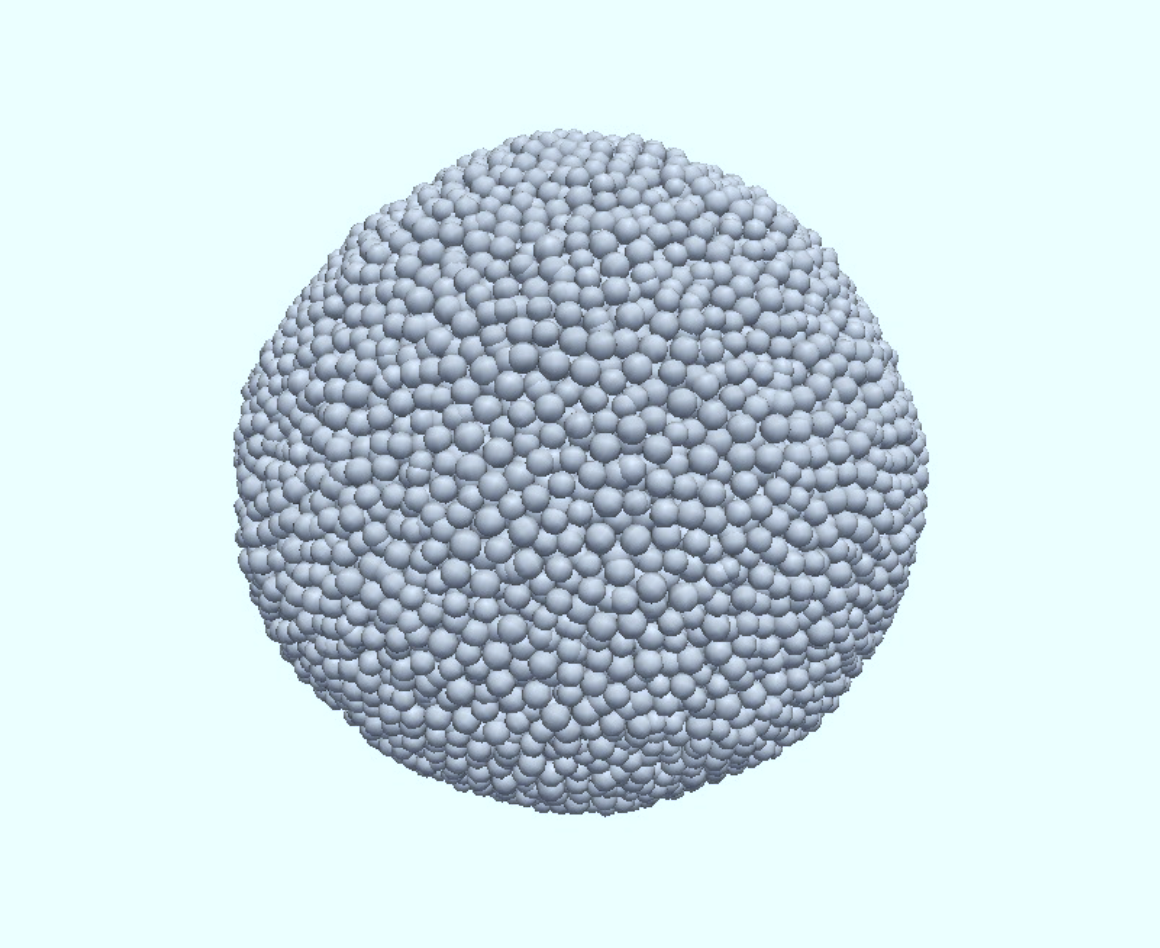} \\
\includegraphics{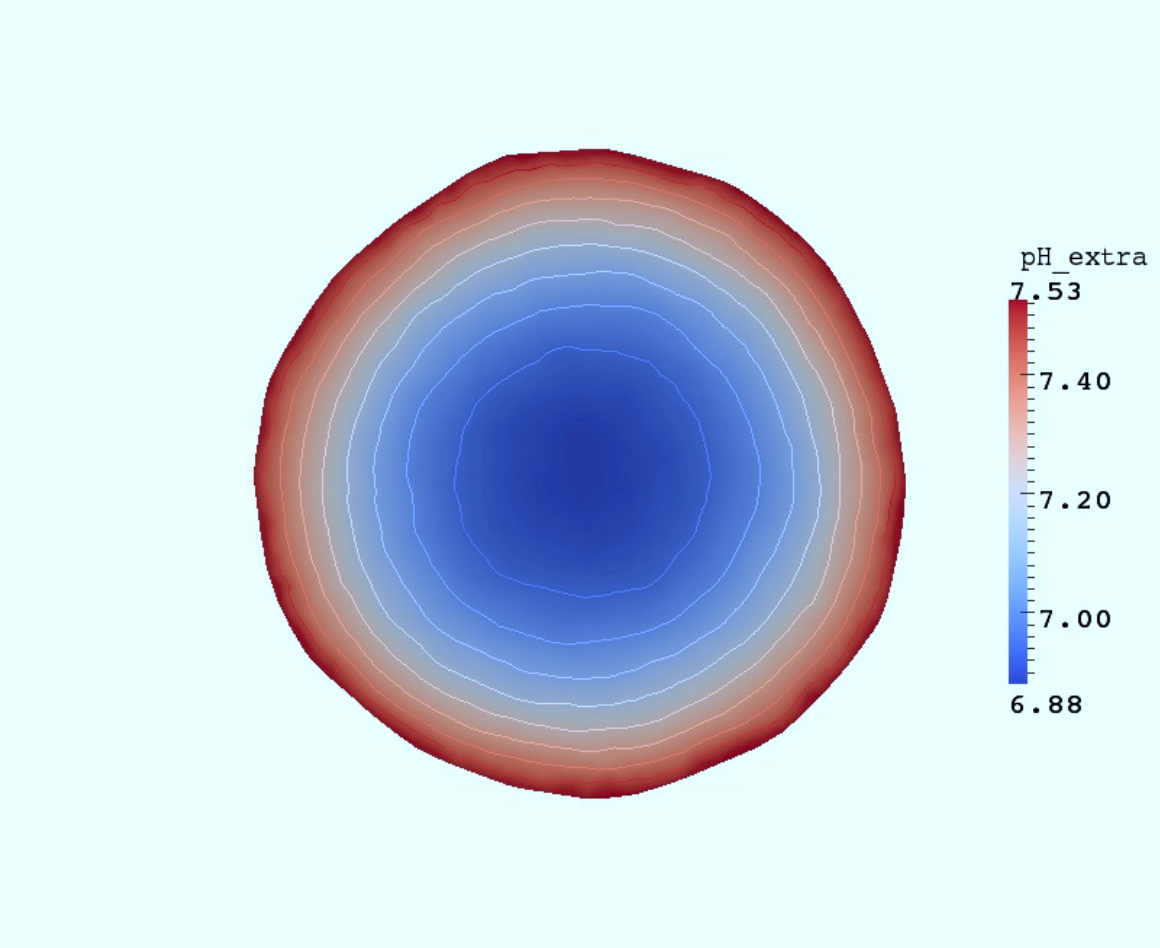} 
\end{array}$
}
\end{center}
\caption{3D picture of small, compact spheroid (upper panel), and a central section showing a color density plot of acidity (pH in extracellular spaces) with superposed contour lines (lower panel). This spheroid contains almost 30000 cells, and these pictures have been produced with ParaView\cite{paraview}.}
\label{small-1}
\end{figure}

\begin{figure}[!ht]
\begin{center}
\includegraphics[width=\linewidth]{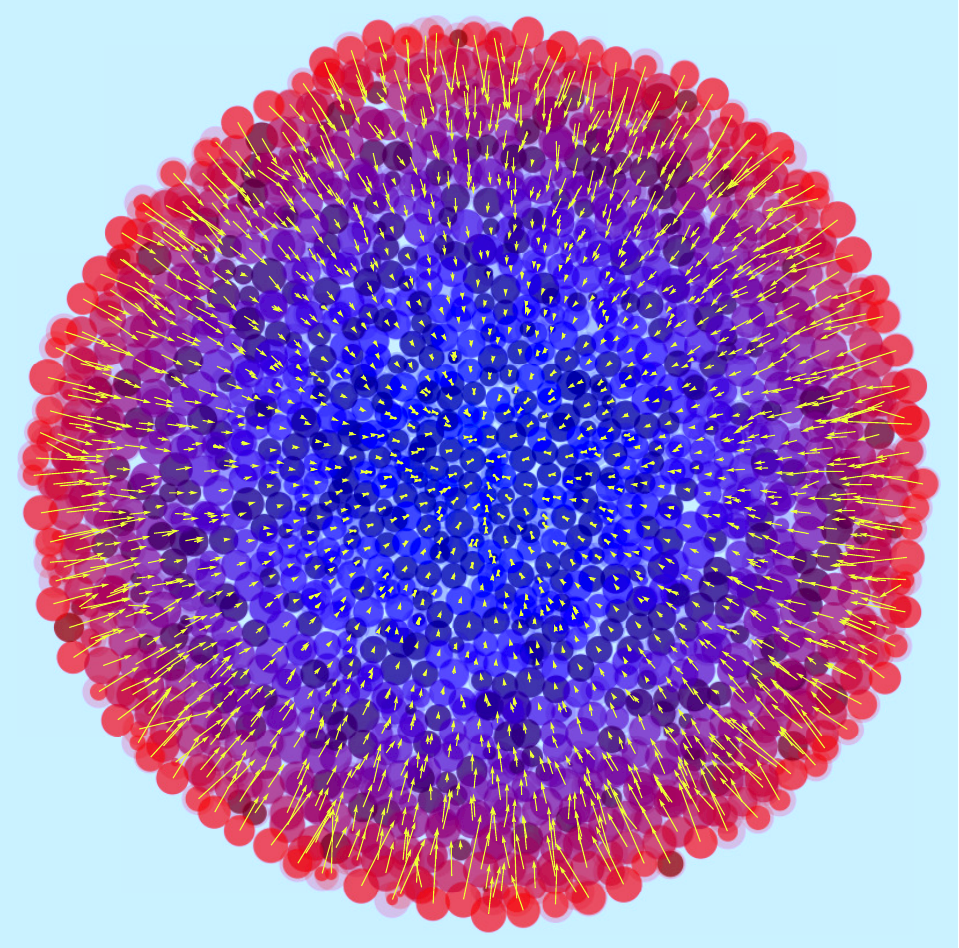}
\end{center}
\caption{Oxygen flow in a central slice of the spheroid of figure \ref{small-1}. This representation has been obtained with {\it Mathematica}\cite{Mathematica8}: the circles represent cross-sections of individual cells, color maps the O$_2$ concentration (red, high concentration; blue, low concentration), the arrow lengths represent the intensity of O$_2$ flow, and their direction is the direction of the flow (arrow length in arbitrary units). The flow is uniformly inward, as cell metabolism consumes O$_2$.}
\label{small-2}
\end{figure}

\begin{figure}[!ht]
\begin{center}
\includegraphics[width=\linewidth]{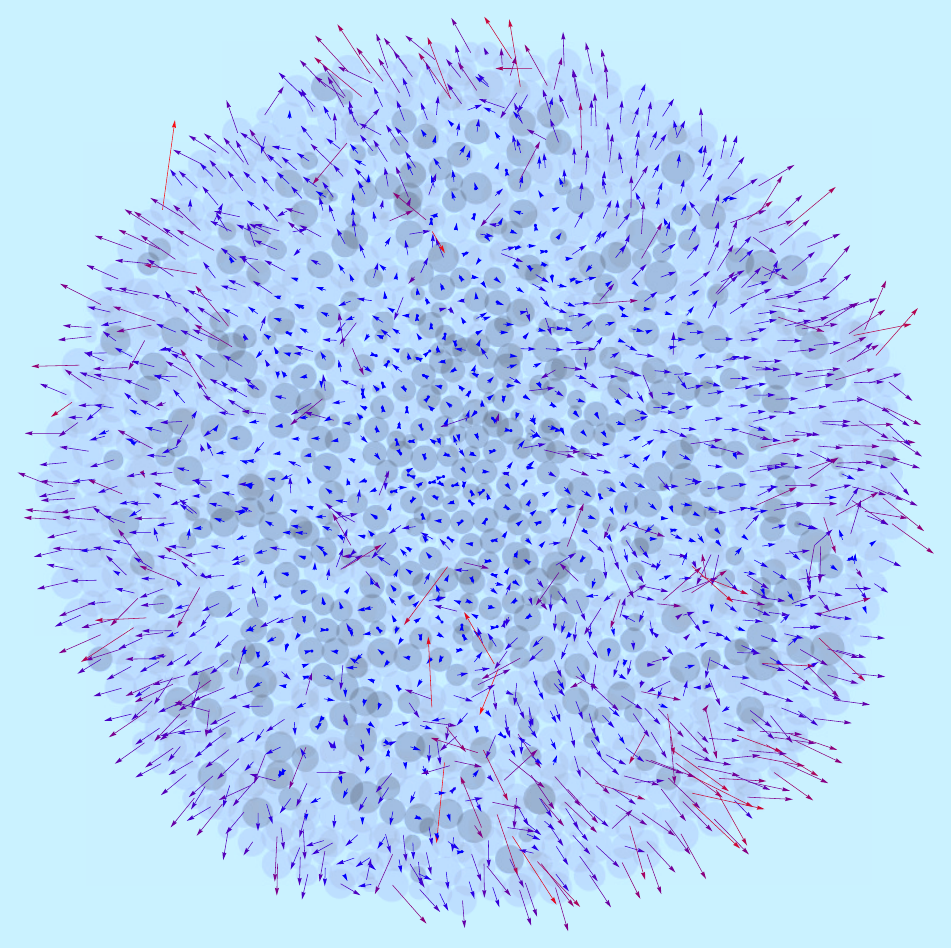}
\end{center}
\caption{Velocity field in the same central slice shown in figure \ref{small-2}. This representation has been obtained with {\it Mathematica}\cite{Mathematica8}: the circles represent again the cross-sections of individual cells, color maps the state of cells (light gray means live cell, dark gray means dead cell), the arrow lengths represent speed, and their direction is the direction of the velocity vector (arrow length in arbitrary units). The velocity field is uniformly outward in the outer shell of viable cells, while it becomes chaotic in the core, where the activity of a few live cells mixes with the shrinking of dead cells. This pattern becomes much more evident in large tumor spheroids\cite{milotti2010emergent}.}
\label{small-3}
\end{figure}

\begin{figure}[!ht]
\begin{center}
\includegraphics[width=\linewidth]{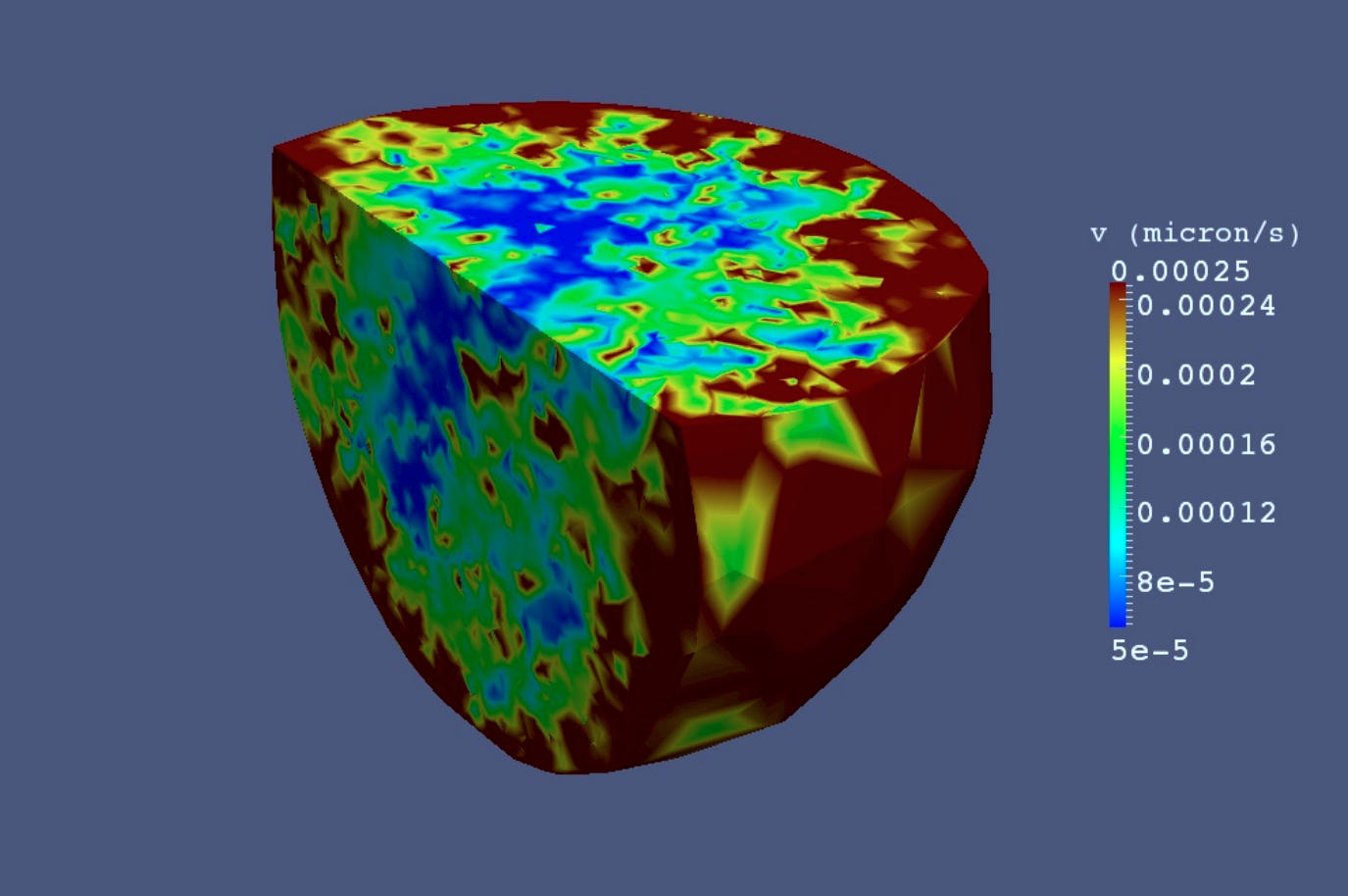}
\end{center}
\caption{Magnitude of the velocity field in  the spheroid of figure \ref{small-1}. Like figure \ref{large3D-Vclipped}, this representation has been obtained with ParaView\cite{paraview} and it should be compared with figure \ref{small-3}, which displays the same velocity field, in vector form and in a single slice: these representations carry slightly different informations and are complementary to each other.}
\label{small-4}
\end{figure}

\begin{figure}[!ht]
\begin{center}
\resizebox{4in}{!}{%
$\begin{array}{c}
\includegraphics{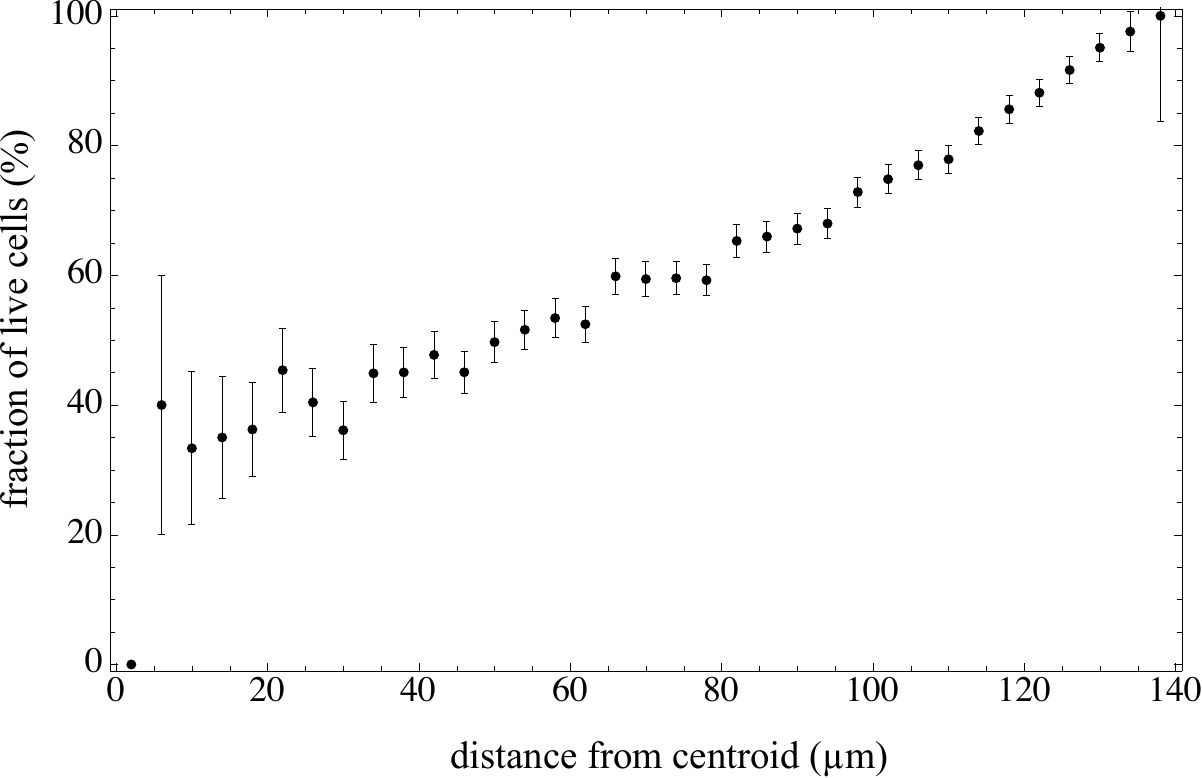} \\
\\
\\
\\
\\
\\
\includegraphics{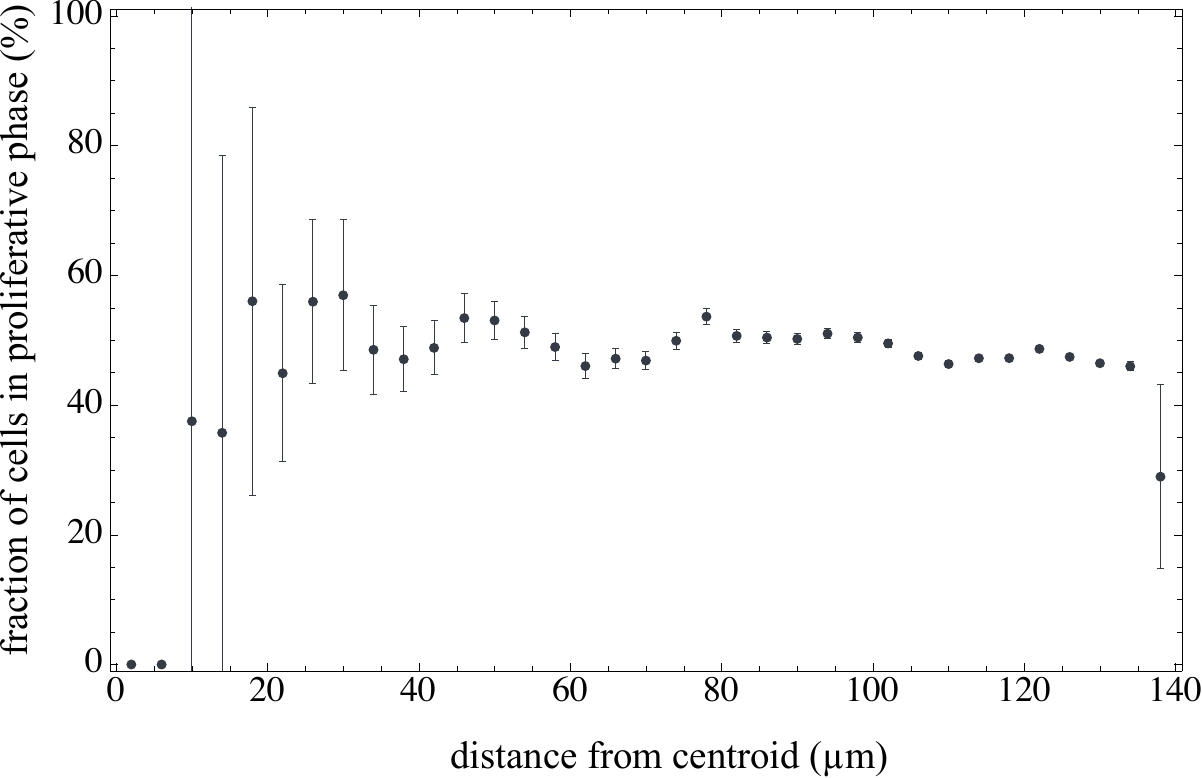} 
\end{array}$
}
\end{center}
\caption{Conventional representations for the small spheroid of figures \ref{small-1}-\ref{small-4}. The upper panel shows the percent fraction of live cells vs. the distance from the centroid ($\mu$m): this plot is unsurprising, it tells us that the deeper we go into the spheroid, the lower the fraction of live cells. The error bars are computed assuming that individual samples have a Poisson statistics, and the large fluctuations at large distance depend on the deviations from exact sphericity. The lower panel shows the percent fraction of live cells that are in a proliferative phase (S+G2+M) vs. the distance from the centroid ($\mu$m): this plot is much more interesting, because it shows some distinct oscillations that indicate depth-dependent synchronization of cell subpopulation. This behavior can justify a heterogenous response to drugs which have a phase-dependent action.}
\label{small-5}
\end{figure}

\section{Conclusions \label{concl} }

We started this paper with some very general considerations, and then we delved into the many details that must be taken into account to carry out even a minimal simulation of the machinery of real tumors. Now we wish to conclude stressing two important and correlated features of the simulation program that make it both extendible and very general, and secure it firmly to experiments: modularity, and the choice of parameter values. 

Modularity is one of the of the main features of the simulation program. This means that the basic procedure for implementing diffusion and transport is nearly the same for most substances, and that inclusion of additional molecular circuits is almost straightforward. The real difficulties with any such addition are twofold: first we have to determine the actual action of the molecules in cells and describe it with a differential system, then we have to search the literature or perform experiments to fix all the model parameters. If we are lucky, these submodels may already exist, however it often happens that either they do not exists, or they are unsuitable for inclusion in the program. There are many reasons for this unsuitability, e.g., the models may give a qualitative description of molecular action, but they have an inordinate number of unknown parameters. In such cases we have to replace the particular model with a phenomenological description with  a greatly reduced parameter set, and determine the new parameters from existing data. Indeed, while developing the program, we have put a lot of work in the determination of parameter values: thus, although the present version of the program has on the whole about one hundred parameters, they are nearly fixed, with very little variability. 

This attention to actual parameter values produces a close connection with experiments, and it is in this sense that we believe that our modeling efforts are more realistic than other similar attempts. Indeed there are several important details that qualify our simulation program as ``realistic''. Firstly, although a truly realistic {\it ab initio} simulation would start from the molecular level, the present choice is only a step higher in the ladder of complexity, as we model single molecular circuits, with experimentally derived parameters. Then, we note that the biochemical modularity means that we can improve the accuracy of simulations simply by increasing the number of biochemical paths that are included in the program. We also stress that although the biomechanical part is heavily parameterized, this is neither a lattice model nor  a cellular automaton, and thus it does not bear the burden of some artificially imposed symmetry in cellular motion. Moreover, the model is not just qualitative, but quantitative as well, since parameters are nearly fixed, and thus it is predictive and can be falsified by measurements. Thus measurements actually drive the development of the simulation program, as in a standard physical framework. Finally,  the basic equations of the model are biophysically and biochemically well motivated, and make it in essence a truly mathematical model. And yet it offers both a valid computational scheme and clearly displays the limitations of present-day computers, as it stresses to their limit both memory and speed resources. Thus this computational scheme is in itself an interesting example of multiscale biological model, which may eventually help contribute to the current debate on computability issues in biology (see, e.g., Refs. \onlinecite{fisher2007executable,hunt2008dichotomies,Hunt:2009:Pharm-Res:19756975}).

\begin{acknowledgments}
The authors acknowledge financial support from the Italian Institute for Nuclear Physics (INFN), and grants from the italian HPC centers CASPUR (Rome) and CINECA (Bologna). 
\end{acknowledgments}


\nocite{*}

%

\end{document}